\documentclass[twocolumn, amssymb,amsmath, aps, prb, showpacs, 10pt]{revtex4-2}
\usepackage{graphicx}
\usepackage{color}
\usepackage{multirow}
\usepackage{hyperref} 
\usepackage{amsmath}
\usepackage{amssymb}
\usepackage{epsfig}
\usepackage{steinmetz}
\DeclareMathAlphabet{\pazocal}{OMS}{zplm}{m}{n}
\usepackage{graphicx}
\usepackage{color}
\usepackage{bm}
\usepackage[normalem]{ulem}
\usepackage{soul}
\usepackage{braket}
\usepackage{csquotes}

\begin{document}
\title {The interplay between structural, magnetic and electronic states in the pyrochlore iridate  Eu$_2$Ir$_2$O$_7$}
\author{Manjil Das$^1$, Sayantika Bhowal$^{1,2}$, Jhuma Sannigrahi$^3$, Abhisek Bandyopadhyay$^4$, Anupam Banerjee$^1$, Giannantonio Cibin$^5$, Dmitry Khalyavin$^6$, Niladri Banerjee$^7$, Devashibhai Adroja$^{6,8}$, Indra Dasgupta$^1$, Subham Majumdar$^1$}
\email{sspsm2@iacs.res.in}

\affiliation{$^1$School of Physical Sciences, Indian Association for the Cultivation of Science, 2A \& B Raja S. C. Mullick Road, Jadavpur, Kolkata 700 032, India}

\affiliation{$^2$Materials Theory, ETH Zurich, Wolfgang-Pauli-Strasse 27, 8093 Zurich, Switzerland}

\affiliation{$^3$School of Physical Sciences, Indian Institute of Technology Goa, Farmagudi, Ponda 403401, Goa, India}

\affiliation{$^4$Department of Physics, Indian Institute of Science Education and Research, Pune, Maharashtra 411008, India}

\affiliation{$^5$Diamond Light Source Ltd., Diamond House, Harwell Science and Innovation Campus, Didcot, Oxfordshire OX11 0DE, UK}

\affiliation{$^6$ISIS Neutron and Muon Source, Science and Technology Facilities Council, Rutherford Appleton Laboratory, Didcot OX11 0QX, United Kingdom}

\affiliation{$^7$Department of Physics, Loughborough University, Loughborough, Leicestershire LE 11 3TU, United Kingdom}

\affiliation{$^8$Highly Correlated Matter Research Group, Physics Department, University of Johannesburg, P.O. Box 524, Auckland Park 2006, South Africa}

\begin{abstract}
We address the concomitant metal-insulator transition (MIT) and antiferromagnetic ordering in the novel pyrochlore iridate Eu$_2$Ir$_2$O$_7$ by combining x-ray absorption spectroscopy, x-ray and  neutron diffractions  and density functional theory (DFT) based  calculations. The temperature dependent powder x-ray diffraction clearly rules out any change in the lattice symmetry below the MIT, nevertheless a clear anomaly in the Ir-O-Ir bond angle and Ir-O bond length is evident at the onset of MIT. From the x-ray absorption near edge structure (XANES) spectroscopic study of Ir-$L_3$ and $L_2$ edges, the effective spin-orbit coupling is found to be  intermediate, at least quite far from the strong atomic spin-orbit coupling limit. Powder neutron diffraction measurement is in line with  an {\it all-in-all-out} magnetic structure of the Ir-tetrahedra in this compound, which is quite common among rare-earth pyrochlore  iridates.  The sharp change in the Ir-O-Ir bond angle around the MIT possibly arises from the exchange striction mechanism, which  favors an enhanced electron correlation via weakening of Ir-Ir orbital overlap and an insulating phase below $T_{MI}$. The theoretical calculations  indicate an insulating state for shorter bond angle validating the experimental observation. Our DFT calculations show a possibility of  intriguing topological phase below a critical value of the Ir-O distance, which is shorter than the experimentally observed bond length. Therefore, a topological state may be realized in bulk Eu$_2$Ir$_2$O$_7$ sample if the Ir-O bond length can be reduced by the application of sufficient external pressure.
\end{abstract}

\maketitle

\section{Introduction}

Recent studies on the 5$d$/4$d$ electron systems such as iridium based pyrochlore oxides have generated significant interest due to the presence of strong spin-orbit coupling (SOC) as well as frustrated lattice geometry. They are in the forefront of exciting new physics as they accommodate a variety of unconventional quantum mechanical states of matter (e.g. quantum spin liquid, metal-insulator transition, Weyl semimetals, topological insulators, non-Fermi liquids, magnetic monopole-like phase, etc.)~\cite{wan,weyl2,matsu,ishi,anupam,monopole} by a unique combination of extended 5$d$/4$d$ orbitals, electron correlations, small magnetic moments, geometric frustration, noncubic crystal field, single ion anisotropy and spin-orbit interaction effects~\cite{Clancy-prb2016}.
\par
The concomitant occurrence of metal-insulator transition (MIT) at the onset of {\it all-in-all-out} (AIAO) long-range antiferromagnetic (AFM) order of the Ir$^{4+}$-tetrahedra is one of the most fascinating aspects of $R_2$Ir$_2$O$_7$ (R = Y and Nd-Lu) pyrochlore family of iridates~\cite{naturecom2015,prl2015,tomiyasu,sagayama,disseler,matsuhira}. In the AIAO structure, spins at the four corners of each Ir tetrahedron point directly inward (all-in) or outward (all-out), {\it i.e.}, toward or away from the center of the tetrahedron~\cite{naturecom2015,prl2015,tomiyasu,sagayama,disseler,matsuhira,shapiro,disseler2,yang1,matsu}. The spins on each neighboring tetrahedron align in the opposite direction. Interestingly,  $R$ = Nd-Lu compounds exhibit metallic behavior at high temperatures followed by MIT upon lowering the temperature ~\cite{matsu,matsuhira,radii}. Only  Pr$_2$Ir$_2$O$_7$ does not possess either MIT or long-range AFM ordering, rather offering quantum spin liquid ground state~\cite{pr2ir2o7}. The MIT must be connected to the magnetic ordering of Ir-sublattice, as the effect is also observed for non-magnetic rare-earth such as Lu. 
\begin{figure}[t]
	\centering
	\includegraphics[width = 8 cm]{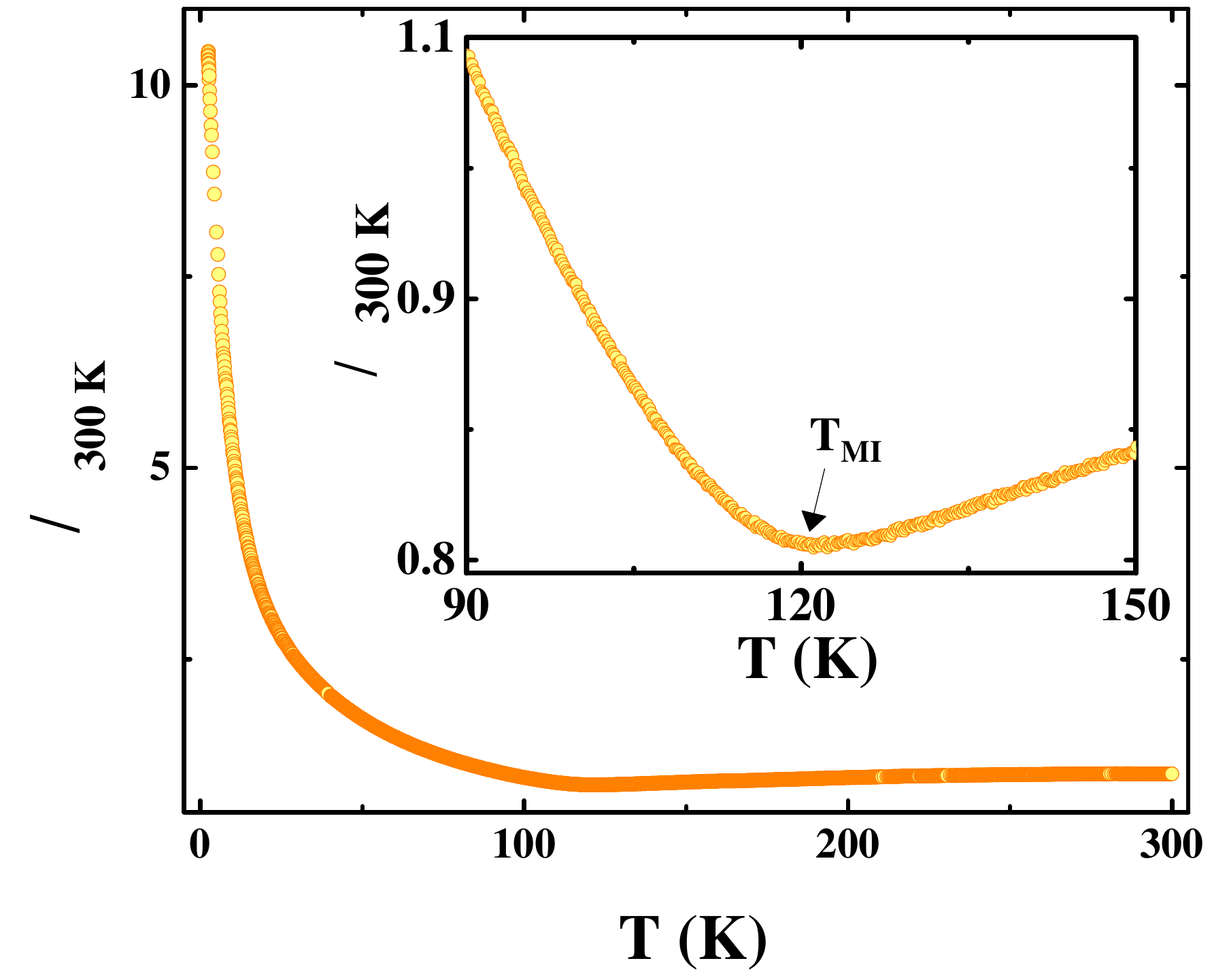}
	\vskip 1 cm
	\caption{The variation of electrical resistivity as a function of temperature. The metal insulator transition temperature is marked by $T_{MI}$.}
	\label{mtrt}
\end{figure}

\par
Eu$_2$Ir$_2$O$_7$ deserves a special mention in this series as it has been predicted to be  a potential topological Weyl semimetal~\cite{pesin,wan,yang1}. In this compound, Eu$^{3+}$ is supposed to be  non-magnetic as the total angular momentum $J$ = 0, while Ir attains 4+ low spin state with the effective total angular momentum $J_{eff}$ = 1/2. Eu$_2$Ir$_2$O$_7$ displays a metal-insulator transition with insulating ground state below $T_{MI}$ = 120 K which coincide with the N\'eel temperature (N\'eel temperature, $T_N$ = $T_{MI}$). Fujita {\it et al}~\cite{fujita} studied the galvanomagnetic properties of single crystalline thin film of Eu$_2$Ir$_2$O$_7$, where they  stabilized the two degenerate  domain structures by the polarity of the cooling magnetic field. Recently, Liu {\it et al}~\cite{liu} reported the Weyl nodes in epitaxial thin film of Eu$_2$Ir$_2$O$_7$ due to the broken cubic symmetry at the bottom and top surfaces of the thin film.
\par
Despite  the wealth of theoretical and experimental works on this rare-earth pyrochlore iridate, the Weyl semimetallic phase in  bulk Eu$_2$Ir$_2$O$_7$ is still elusive. On the other hand, the  true origin of MIT as well as the nature of insulating ground state have  not been well understood yet. Raman spectroscopic studies on Eu$_2$Ir$_2$O$_7$ confirmed the presence of strong anomalies in the Ir-O-Ir bending mode, $E_g$, across $T_N$, which may modulate the superexchange or Dzyaloshinskii-Moriya  exchange interaction between  Ir ions~\cite{uda}. Although the bulk susceptibility and $\mu$SR studies indicated broken time-reversal symmetry below $T_{MI}$, very little direct information is available to test the theoretical prediction of the ordered moment direction and periodicity. The main reason behind this is the technical challenge in neutron scattering experiments on iridate compounds, as Ir is a strong neutron absorber. In case of Nd$_2$Ir$_2$O$_7$ and Tb$_2$Ir$_2$O$_7$, both the Ir and the rare-earth moments are predicted to form the AIAO structures~\cite{nd2ir2o7,tb2ir2o7}.  In the neutron diffraction study, the magnetic intensity at low temperature was detected due to the ordering of the rare-earth moments~\cite{tomiyasu}. However, no direct signal for the ordered Ir moments was  detected, probably because of Ir's weak magnetic signal compared to the rare-earth. Recent investigations of $\mu$SR, resonant elastic x-ray scattering, and resonant inelastic x-ray scattering on Eu$_2$Ir$_2$O$_7$ directly confirmed the ordering of Ir magnetic moments with a wave vector $k$ = (0, 0, 0)~\cite{zhao, sagayama,Clancy-prb2016}. These studies argue that the AIAO structure of Ir sublattice is the most probable symmetry allowed spin structure as there is no change in the crystal symmetry across $T_{MI}$. Evidently, a neutron diffraction study on Eu$_2$Ir$_2$O$_7$ is important for an explicit knowledge of the  magnetic structure. 
\par
We employed  in-depth temperature ($T$) dependent synchrotron based high resolution powder x-ray diffraction, x-ray absorption spectroscopy and neutron powder diffraction for precise structural investigations across the MIT.  Density functional theory  (DFT) based {\it ab initio} calculation was  performed to support our results of the magnetic structure and identify the bulk electronic structure.

\begin{figure}[t]
\centering
\includegraphics[width = 9 cm]{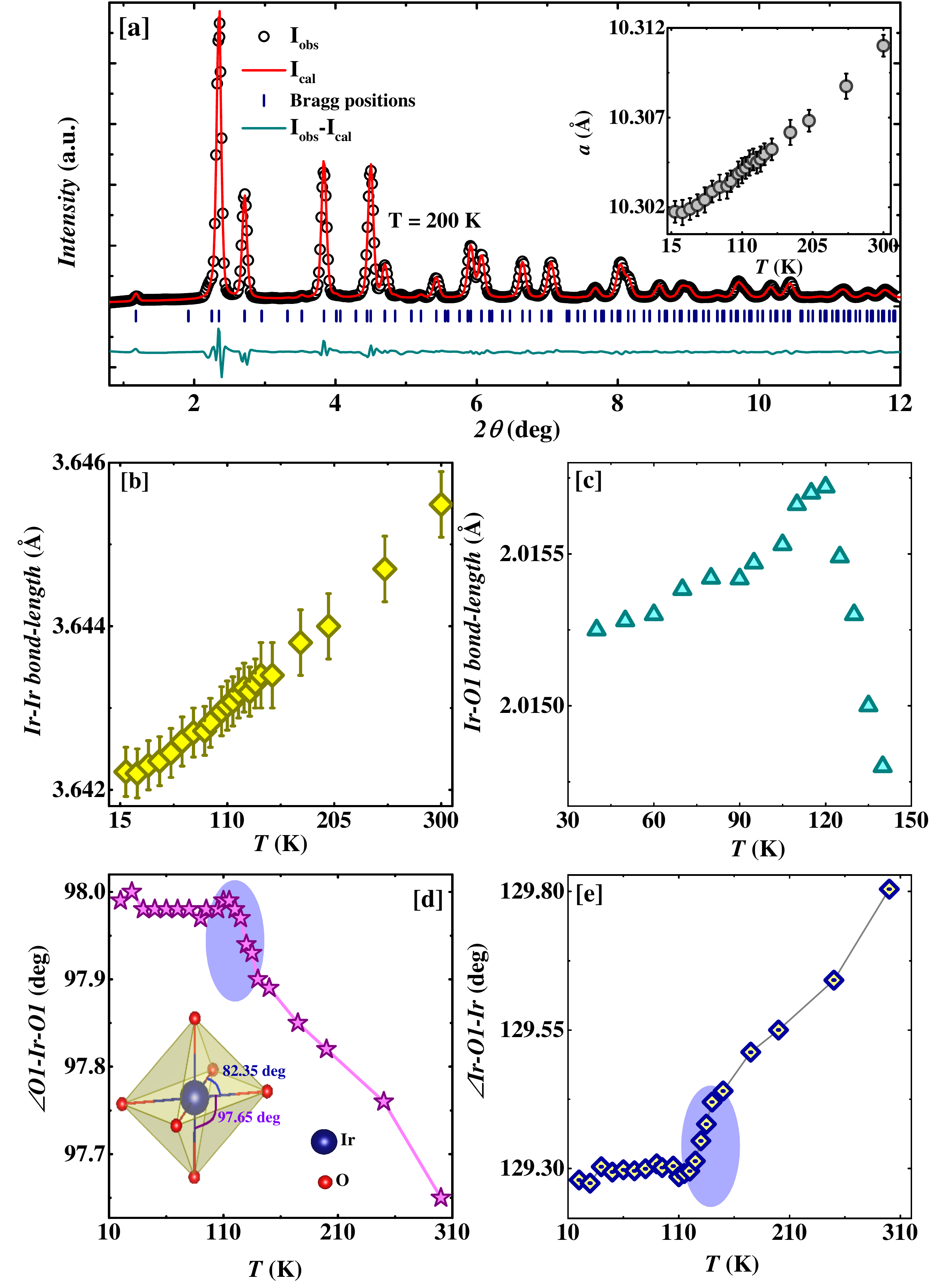}
\caption {[a] Synchrotron based powder x-ray diffraction pattern along with the refinement measured at 200 K. Inset: variation of the cubic  lattice parameter with temperature; [b], [c], [d] and [e] respectively illustrate the temperature variations of Ir-Ir bond length, Ir-O1 bond length, $\angle$(O1-Ir-O1) and $\angle$(Ir-O1-Ir). The inset of [d] shows a single IrO1$_6$ octahedron, where the deviation of $\angle$(O1-Ir-O1) from 90$^{\circ}$ is evident.}
\label{pxrd}
\end{figure}

\begin{figure}[t]
\centering
\includegraphics[width = 8 cm]{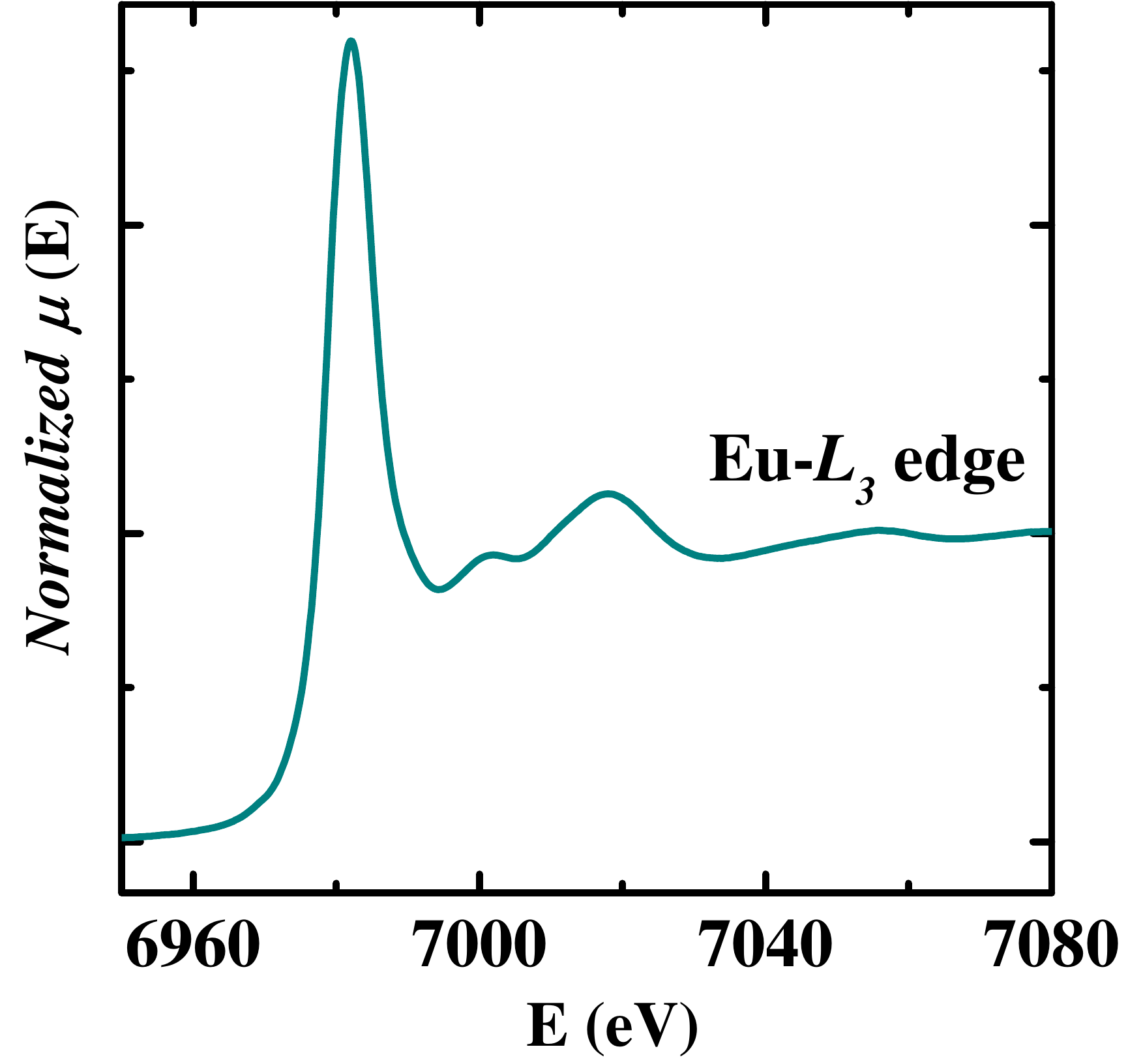}
\vskip 1 cm
\caption{XANES spectra of the Eu-$L_3$ edge of Eu$_2$Ir$_2$O$_7$ at 200 K.}
\label{Eu-xanes}
\end{figure}

\begin{figure*}[t]
\centering
\includegraphics[width = 14 cm]{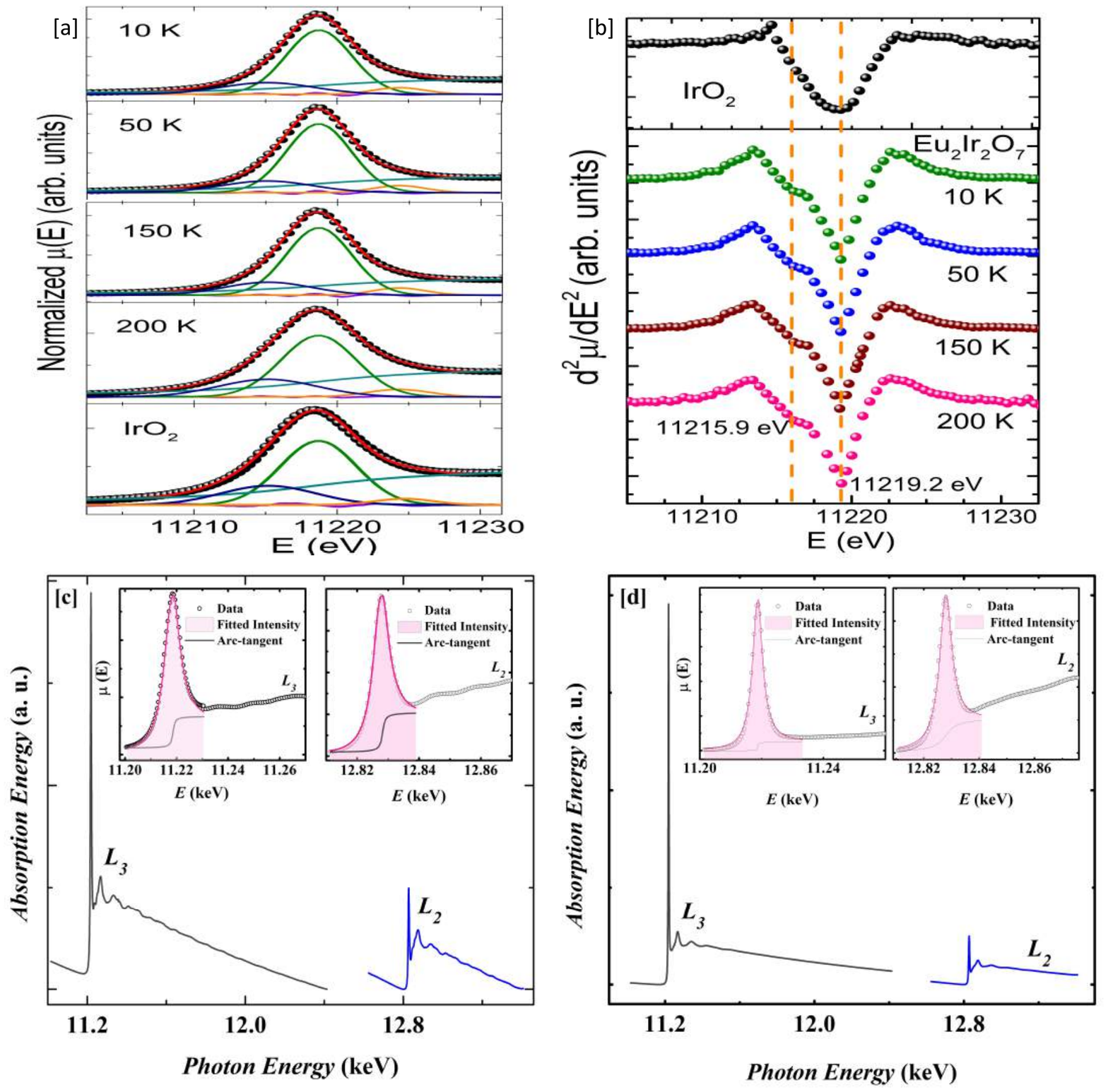}
\vskip 1 cm
\caption{The measured Ir-$L_3$ edge XANES spectra of Eu$_2$Ir$_2$O$_7$ for the four selected temperatures along with the respective fitted curves are shown in [a]. For comparison, the Ir-$L_3$ edge XANES data together with fitting for the reference IrO$_2$ sample have also been represented in the same panel. XANES spectra of the Ir-$L_3$-edge of Eu$_2$Ir$_2$O$_7$ at four different temperatures along with that of IrO$_2$ are represented. [b] shows the second derivative curves of Ir-$L_3$ edge of Eu$_2$Ir$_2$O$_7$ and IrO$_2$ . [c] and [d] illustrate Ir-$L_3$ and $L_2$ edge XANES spectra measured at 200 K for Eu$_2$Ir$_2$O$_7$ and IrO$_2$ respectively. The fitted Ir-$L_3$ and Ir-$L_2$ XANES data using a combination of  Lorentzian peak and arc tangent background function are presented in respective insets.}
\label{Ir-xanes}
\end{figure*}

\section{Methodology}
\subsection{Experimental techniques}
Polycrystalline sample of Eu$_2$Ir$_2$O$_7$ was prepared through solid state reaction technique using high purity ($>$99.9\%, Sigma Aldrich) Eu$_2$O$_3$ and IrO$_2$ as starting materials~\cite{anupam}. The phase purity and crystal structure of the synthesized sample were initially confirmed by the laboratory-based powder x-ray diffraction (XRD) using Cu $K_{\alpha}$ radiation. The temperature dependent powder x-ray diffraction (PXRD) measurements were carried out  at the P21.1 beamline of PETRA III Synchrotron radiation facility, Germany, using the monochromatic x-ray of 0.122 \AA~ wavelength. The PXRD data were refined using the MAUD software package for the Rietveld refinements~\cite{maud}.  X-ray absorption spectroscopy (XAS) measurements were carried out at the B-18 beamline, Diamond Light Source, UK, around the Eu-$L_3$ edge  at 200 K, as well as the Ir-$L_3$ and $L_2$ edges  at different temperatures in standard transmission geometry. X-ray absorption near edge structure (XANES) and extended x-ray absorption fine structure (EXAFS) parts of the XAS data  were analyzed with freely available DMETER software packages~\cite{dem1,dem2}, applying a multishell data refinement procedure~\cite{VMo-DP-PRB,srimanta-prb}. The analysis were  implemented considering average Eu and Ir local atomic structure model calculated from the XRD Rietveld refinement results.
\par
Neutron diffraction experiment was performed on powder sample of Eu$_2$Ir$_2$O$_7$ at the ISIS facility (UK) where high-resolution data were collected on heating at 1.5 K and 150 K at the WISH diffractometer~\cite{isis}. The powder sample of Eu$_2$Ir$_2$O$_7$ (weighing about 3 g) was placed in a thin hollow shaped cylinder in order to mitigate the effect of the strong neutron absorption of Eu and  Ir nuclei as much as possible. The neutron data were acquired at two temperatures (1.5 and 150 K) for long measurement time (approximately 24 hours at each temperature) for better statistics. Rietveld refinements were carried out on the collected neutron powder diffraction data using the FULLPROF program~\cite{Fullprof}.
\par
The resistivity $\rho$ was measured using [see fig.~\ref{mtrt}] the four-probe method on a cryogen-free  system (Cryogenic Ltd.). The sample shows clear signature of metal insulator transition at around 120 K, which matches well with the previous reports~\cite{anupam}. The sample was also investigated through core-level x-ray photoelectron spectroscopy (XPS) at room temperature using Al $K_{\alpha}$ radiation on a laboratory-based commercial instrument (Omicron) with in situ surface cleaning by argon ion sputtering.

\subsection{Theoretical methods}
The basic electronic structure and the MIT in Eu$_2$Ir$_2$O$_7$ were studied using first principles calculations based on DFT. To determine the magnetic ground state, we computed the total energies of the various symmetry allowed magnetic structures using the plane-wave based projector augmented wave (PAW)~\cite{PAW1,PAW2} method as implemented in the Vienna {\it ab initio} simulation package (VASP)~\cite{vasp1,vasp2} within the generalized gradient approximation (GGA) including Hubbard $U$~\cite{U}. The kinetic energy cut off for the plane wave basis was chosen to be 550 eV and a $\Gamma$ centered 4$\times$4$\times$4 k-mesh was used for the Brillouin-Zone integration. The Eu-$f$ states were kept frozen at the core~\cite{fc} and an effective Hubbard interaction $U_{\rm eff} = U-J$ = 2-0.5 = 1.5 eV was considered at the Ir site.  For the calculation of topological properties in the small Ir-O bond length regime, we extract the low energy tight-binding  model  Hamiltonian  of the primitive unit cell using maximally localized Wannier functions (MLWFs) formulation \cite{MarzariVanderbilt1997,Souza2001} as implemented in the \texttt{Wannier90} code \cite{Mostofi2008}. With this low energy model, we then compute the $Z_2$ invariant and the surface states using \texttt{WannierTools} \cite{Wu2018}. For the surface state calculations we used a slab of 20 unit-cell thickness stacked along the [110] direction.

\section{Experimental Results}
\begin{figure}[t]
	\centering
	\includegraphics[width = 8 cm]{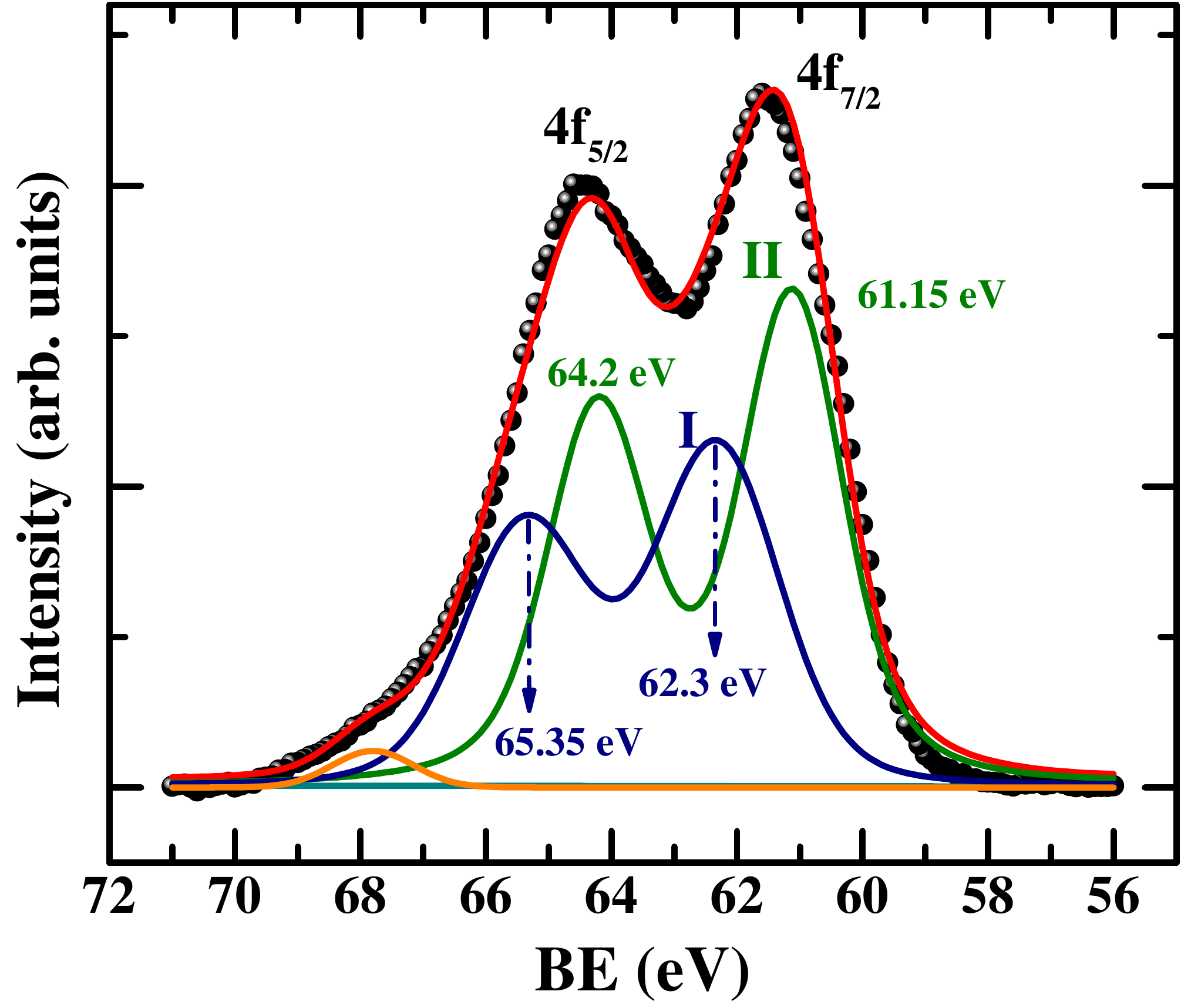}
	\caption{The core-level Ir-$4f$ x-ray photoemission spectroscopy (XPS) study of Eu$_2$Ir$_2$O$_7$ at 300 K is depicted.}
	\label{xps}
\end{figure}

\subsection{Powder x-ray diffraction}
Rietveld refinement of the collected PXRD pattern (fig. \ref{pxrd} [a]) at 200 K ensures a single phase sample of Eu$_2$Ir$_2$O$_7$, crystallizing in the cubic pyrochlore structure with $Fd\bar{3}m$ space group. The crystal structure has two inequivalent oxygen sites, namely O1 ($x$ = $\xi_1$, $y$ = 0.125, $z$=0.125) and O2 ($x$ = 0.125, $y$ = 0.125, $z$ = 0.125).  Each Ir cation forms strongly trigonally distorted IrO1$_6$ octahedral unit, with each single IrO1$_6$ unit being constituted of the acute and obtuse O1-Ir-O1 bond angles of 82.35$^{\circ}$ and 97.65$^{\circ}$ respectively (see inset of fig.~\ref{pxrd}[d]). The extent of such trigonal distortion gradually increases with decreasing temperature till about $T_{MI}$, as illustrated in terms of the deviation of $\angle$(O1-Ir-O1) from perfect cubic 90$^{\circ}$ value. However, the temperature dependent PXRD data do not show any change in the crystal symmetry across the MIT. The cubic lattice parameter $a$ and the Ir-Ir bond length decrease monotonically with decreasing $T$ without any noticeable feature at $T_{MI}$ [see inset of fig.~\ref{pxrd}[a] and fig.~\ref{pxrd}[b] respectively]. Nonetheless, significant anomaly arises in the temperature variations of Ir-O1 bond length and $\angle$(Ir-O1-Ir), as shown in figs.~\ref{pxrd}[c] and [e] respectively. As evident, $\angle$(Ir-O1-Ir) shows a sharp drop at $T_{MI}$ (fig.~\ref{pxrd}[e]), while Ir-O1 bond length shows a peak like feature at around $T_{MI}$.

\begin{figure*}[t]
	\centering
	\includegraphics[width = 14 cm]{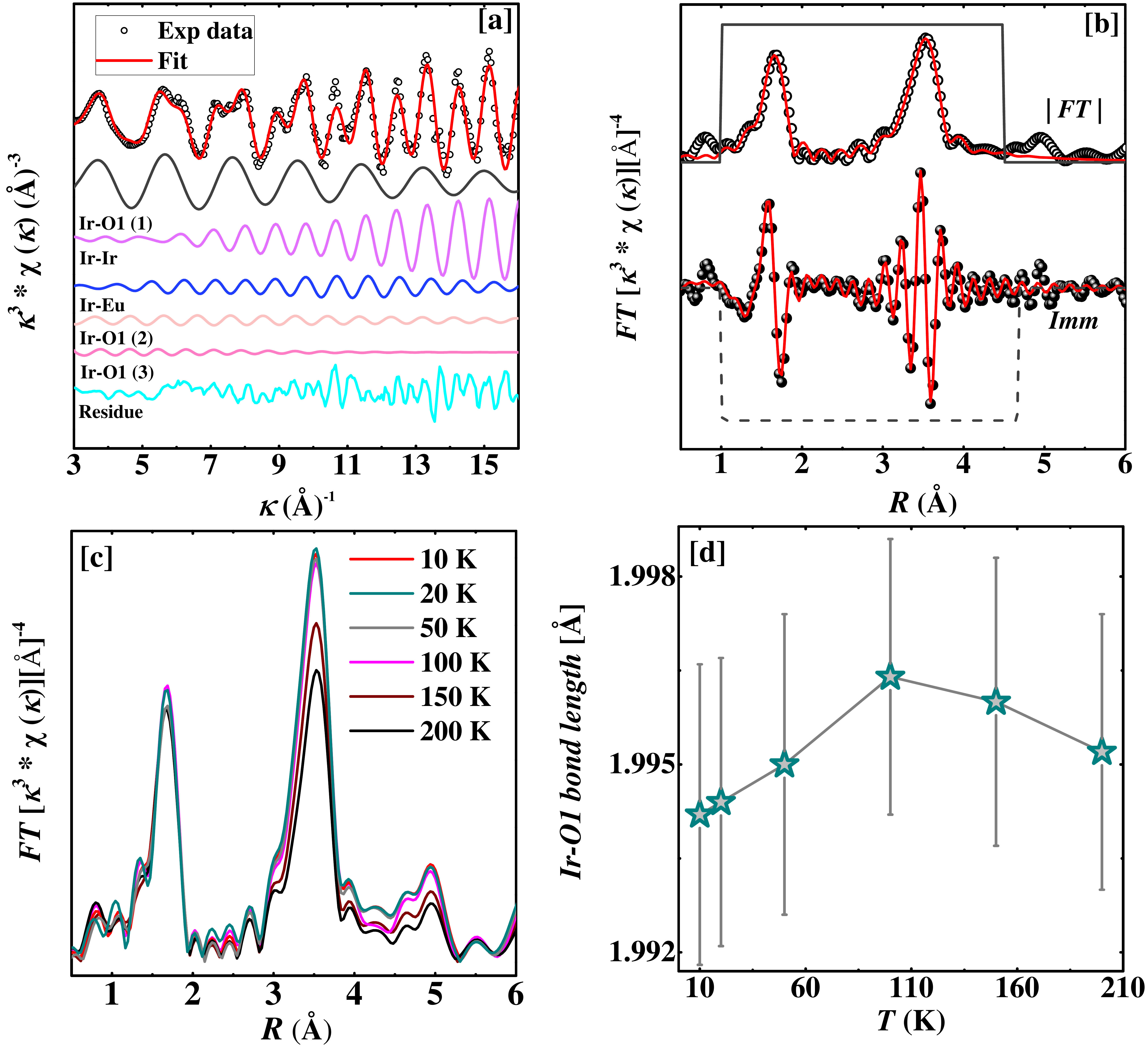}
	\caption{Ir-$L_3$ edge $\kappa^3$ weighted experimental EXAFS data (shaded black circles) at $T$ = 200 K and the corresponding best fit (red solid line) are depicted in [a]; the contributions from the individual single and multiple scattering paths (solid colored line) and the residual [$\kappa^2$$\chi_{exp}$ - $\kappa^2$$\chi_{th}$] (solid cyan line) are also shown, vertically shifted for clarity; [b] the Fourier transforms of the respective experimental data (black circles) and the theoretical fitted curve (solid red line); the magnitude ($|FT|$) and the imaginary parts ($Imm$) are indicated; vertically shifted for clarity. [c] denote the FT data at different temperatures as a function of $R$. [d] The variation of Ir-O1 bond length  with temperature obtained from the Ir-$L_3$ EXAFS analysis.}
	\label{Ir-exafs}
\end{figure*}

\subsection {X-ray absorption near edge structure}
In the stoichiometric Eu$_2$Ir$_2$O$_7$,  Eu and Ir should be ideally in the 3+ and 4+ oxidation states respectively. However, the sample can possess  Eu and Ir in multiple oxidation states due to antisite disorder or oxygen vacancy. To determine the charge states of the cations, we have analyzed the XANES spectra of Eu and Ir $L_3$ edges as shown in figs.~\ref{Eu-xanes},\ref{Ir-xanes} [c].  The energy position ($\sim$6982.3 eV) of the white line of 200 K Eu-$L_3$ XANES spectrum (fig.~\ref{Eu-xanes}) closely matches with that of Eu$_2$O$_3$ ($\sim$ 6983 eV), thus confirming the expected 3+ valence of Eu. Furthermore, a symmetric nature of the XANES peak and the absence of any shoulder-like feature at the low- energy side clearly rule out the possibility of any Eu$^{2+}$ species ($\sim$ 6975.2 eV)~\cite{jpcc,euv} in this sample.

\begin{figure*}[t]
	\centering
	\includegraphics[width = 12 cm]{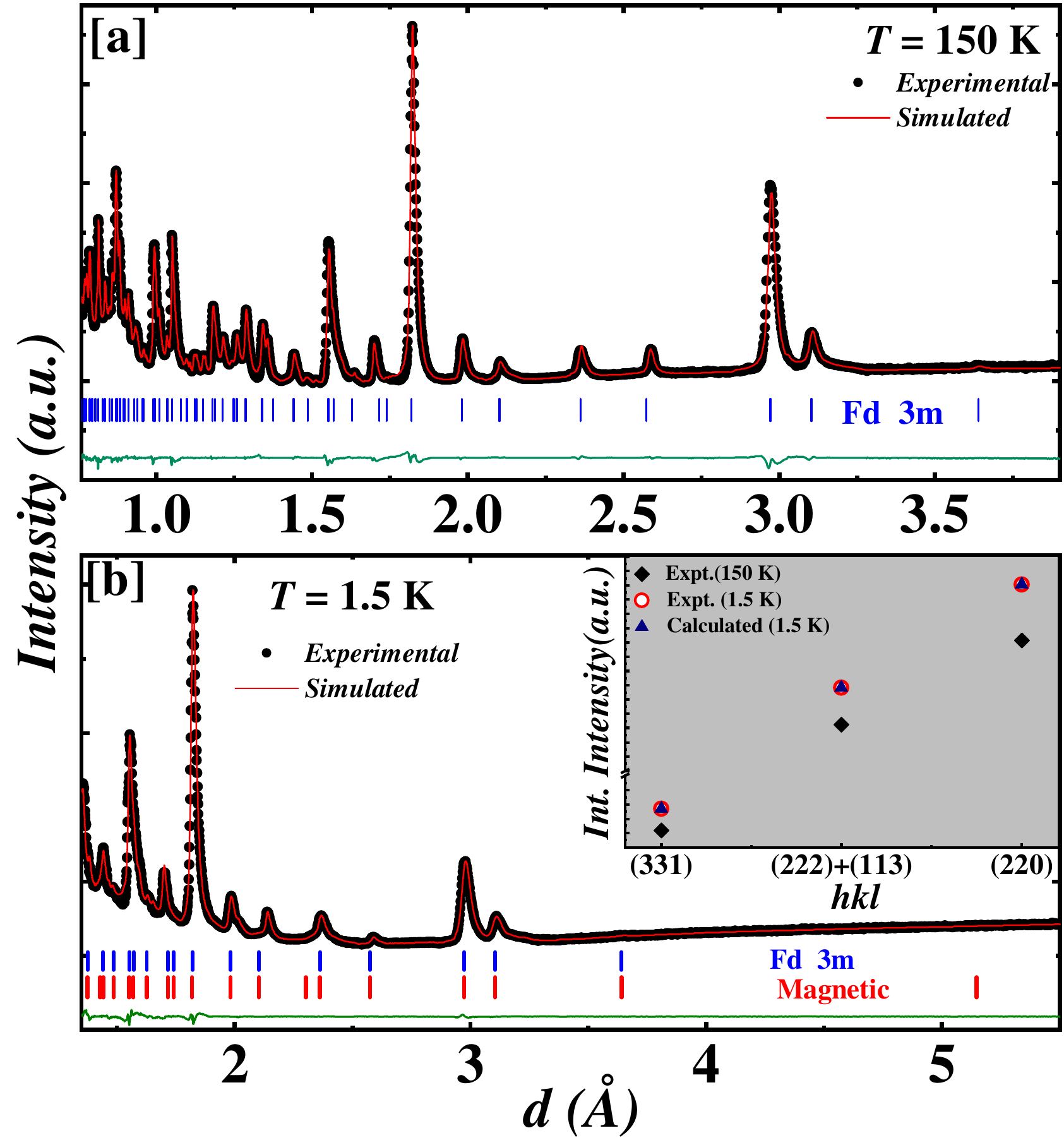}
	\vskip 1 cm
	\caption {[a] and [b]: Powder neutron diffraction pattern recorded at 150  and 1.5 K respectively  along with refinements; inset of [b] shows comparison of the integrated intensity of various Bragg lines recorded at 150 and 1.5 K. Here $d$ denotes the lattice spacing.}
	\label{pnd1}
\end{figure*}

The measured rising edges (2$p$ to 5$d$ transition) of the Ir-$L_3$ XANES spectra for Eu$_2$Ir$_2$O$_7$ and the IrO$_2$ reference have been theoretically fitted (see fig.~\ref{Ir-xanes}[a] ) by fixing the background at the $arctangent$ shape, and the peak width and step at 5.6 eV and 0.9 eV respectively. All the spectra clearly exhibit a weak asymmetry in the form of well resolved low-energy shoulder along with the relatively much intense higher energy peak. The lower energy peak (11215.06 eV) corresponds to Ir 2$p$ $\rightarrow$ $t_{2g}$ transition and the higher energy peak (11218.73 eV) signifies Ir 2$p$ $\rightarrow$ $e_g$ transition. It should be noted that neither the Ir 2$p$ $\rightarrow$ $t_{2g}$ nor the Ir 2$p$ $\rightarrow$ $e_g$ peaks does show any noticeable chemical shift in any of the measured temperatures with respect to the IrO$_2$ spectrum, thus suggesting pure 4+ valence of Ir in this compound~\cite{psmio-prb}. Moreover, the ratio of areas under the Ir 2$p$ $\rightarrow$ $t_{2g}$ and Ir 2$p$ $\rightarrow$ $e_g$ peaks does not reveal any change in the measured spectra of Eu$_2$Ir$_2$O$_7$ relative to the reference IrO$_2$ oxide, hence rejecting any possibility of Ir-valence-change and/or Ir-mixed valency in the Eu$_2$Ir$_2$O$_7$ compound down to the lowest measured temperature. The corresponding second derivative curves (see fig.~\ref{Ir-xanes}[b]), representative of the white line (2$p$ $\rightarrow$ 5$d$ transition) feature, clearly reveal a well-resolved doublet features, corroborating the Ir 2$p$ $\rightarrow$ $t_{2g}$ and Ir 2$p$ $\rightarrow$ $e_g$ transitions~\cite{psmio-prb,choy,choy1}. The peak structures, relative peak intensities and the energy positions corresponding to either of the transitions (see dotted orange lines in fig.~\ref{Ir-xanes}[b]) remain nearly identical for all the temperatures in Eu$_2$Ir$_2$O$_7$, and also match quite well with the findings of previously reported $d^5$ iridates~\cite{psmio-prb,choy}. This further supports the pure 4+ oxidation state of Ir in this compound.

\subsection {X-ray photoelectron spectroscopy}
We further collected Ir-$4f$ spectra and the respective fitting have been shown in fig.~\ref{xps}. Considering the complex nature of the Ir-$4f$ core-lines in case of correlated metallic Ir-oxides~\cite{kahk}, we employ combination of asymmetric $Doniach-Sunjic$ (DS) (for lower BE doublet II) and symmetric $Gaussian-Lorentz$ (for higher BE doublet I) lineshapes during fittings. The background is defined using the Shirley function, while fitting both the lower BE and higher BE spin-orbit doublets (doublet II and doublet I respectively) the area ratio and the spin-orbit energy separation corresponding to each of the doublets were constrained at ~1.3 and ~3.05 eV, respectively~\cite{kennedy}. We assign  the lower BE doublet II as {\it screened} component and the higher BE spin-orbit doublet I as the {\it unscreened} one due to the final state effect following in the literature ~\cite{kotani,kahk,cox,camp}. In addition to these two doublets, a symmetric Lorentz convoluted Gaussian singlet satellite peak (designated by orange color) is also considered at a relatively higher binding energy (~67.8 eV), which can be ascribed to the unscreened Ir-5$p_{1/2}$ conduction electrons~\cite{kahk}. The fit was optimized by allowing the asymmetry parameter and widths ($w$) in the lower BE Doniach-Sunjic doublet II, and the widths of the Gaussian-Lorentz line shape for doublet I to vary freely to achieve satisfactory results. The resulting asymmetry parameter approaches nearly zero, thus transforming the DS lineshape into the standard symmetric Gaussian-Lorentz profile. Finally, the binding energy peak positions of the individual spin-orbit split doublets (4$f_{5/2}$ and 4$f_{7/2}$) together with their respective spin-orbit separation further support the Ir$^{4+}$ valence in our Eu$_2$Ir$_2$O$_7$, in agreement with the findings of Ir-$L_3$ edge XANES results.

\subsection {Branching ratio calculation}
To get a quantitative idea about the Ir-5$d$ SOC in Eu$_2$Ir$_2$O$_7$,~\cite{Laguna-PRL,Cho-JPCM,Clancy-PRB2012,Dwivedi-MRX,Tafti-PRB} as well as to check the influence (if any) of SOC towards the observed MIT, we carefully investigated the Ir-$L_2$ and $L_3$ edge white line intensities. The intensities of the $L_2$ and $L_3$ edge white lines are dominated by the dipolar transition probabilities: 2$p_{1/2} \rightarrow$ 5$d_{3/2}$ and 2$p_{3/2} \rightarrow$ 5$d_{5/2}$,5$d_{3/2}$ respectively~\cite{Cho-JPCM,qi}. The ratio of these two intensities is designated as `Branching Ratio', BR=$I(L_3)/I(L_2)$, which is related to the expectation value of the spin-orbit operator, $<{\bf L.S}>$ via BR =$\frac{(2+r)}{(1-r)}$, where $r = <{\bf L.S}>/n_h$ and $n_h$ is the average number of holes in the 5$d$ state~\cite{Laguna-PRL}. On the other hand, the spin-orbit coupling Hamiltonian is given by, ${\mathcal{H}}_{SOC} = \lambda{\bf L.S}$, where $\lambda$ is the SOC constant. When the effective SOC is negligible, the value of BR approaches 2, and it is called the statistical BR. A value of BR considerably higher than 2, suggests  strong coupling between the local orbital and spin moments~\cite{Laguna-PRL}. The prevailing solid state and electronic factors (e.g. crystal distortion, electronic bandwidth, hopping, hybridization, superexchange, etc) together can renormalize the SOC strength in a solid, and hence the overlap between the spin-orbit-derived rearranged $J_{eff}$ states can result in significantly reduced $<{\bf L.S}>$~\cite{Clancy-PRB2012,Aczel-PRB2019}.
\begin{figure*}[t]
	\centering
	\includegraphics[width=14 cm]{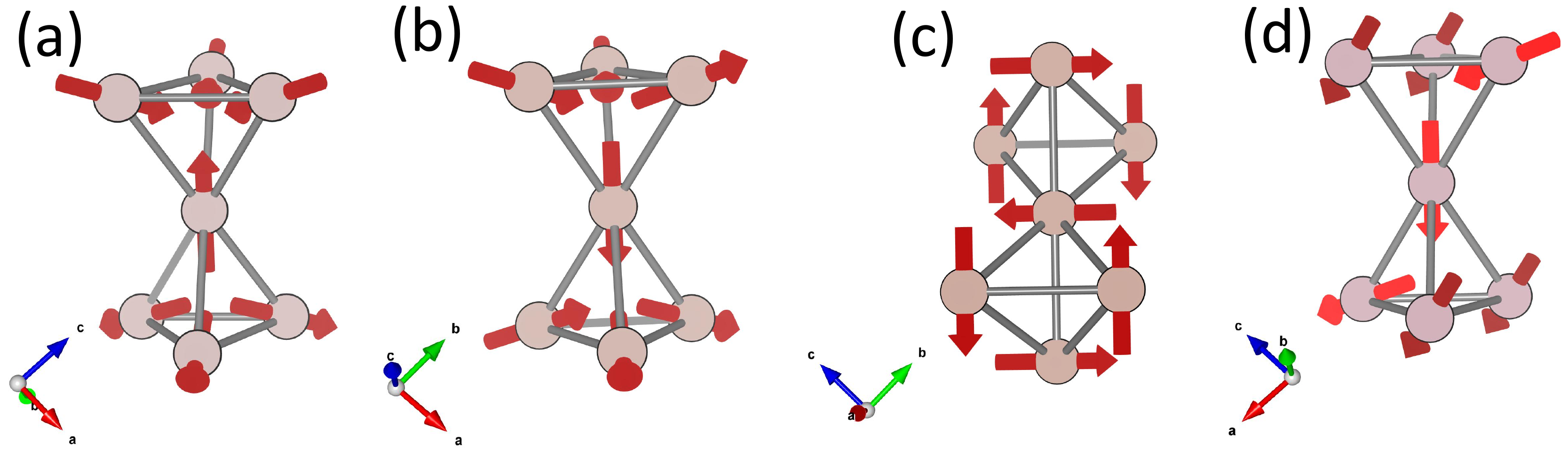}
	\caption{Symmetry allowed magnetic structures of Eu$_2$Ir$_2$O$_7$. The spin configurations at Ir sites corresponding to the magnetic space groups [a]  $Fd\bar{3}m^\prime$, [b]  $I4_1/am^\prime d^\prime$ , [c]  $I4_1^\prime/amd^\prime$, and [d]  $Imm^\prime a^\prime$. The $Fd\bar{3}m^\prime$ magnetic structure, showing the AIAO magnetic configuration, constitutes the ground state of Eu$_2$Ir$_2$O$_7$. The magnetic space group  $Imm^\prime a^\prime$ allows for two different moment values at the Ir site as indicated in two different colors in [d]. The magnetic moments for these two magnetically inequivalent Ir atoms are listed in Table \ref{energy}.}
	\label{spin}
\end{figure*}

\begin{figure*}[t]
	\centering
	\includegraphics[width=14 cm]{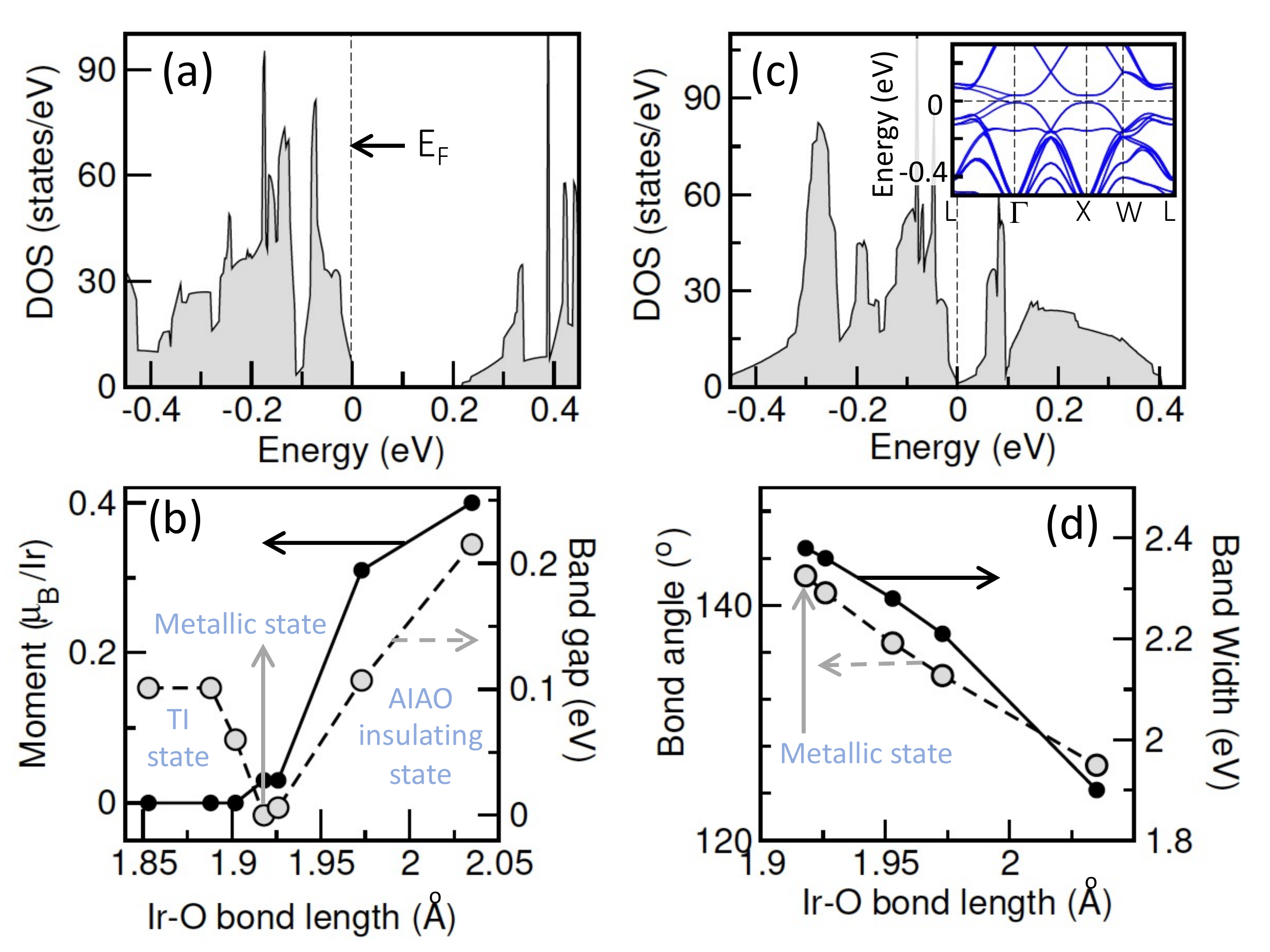}
	\caption{ Metal-insulator transition and the roles of Ir-O bond length and $\angle$Ir-O-Ir bond angle. [a] The total density of states in Eu$_2$Ir$_2$O$_7$ for the AIAO magnetic structure, computed within GGA+SOC+$U$, indicating an insulating ground state. The vertical dashed line indicates the Fermi energy ($\rm {E_F}$). [b] The phase diagram showing the variations of magnetic and electronic properties of Eu$_2$Ir$_2$O$_7$ as a function of Ir-O bond length. The variations of the spin moment and band gap are shown in  solid and dashed lines respectively, indicating a transition from AIAO insulating state to a TI insulating state via a metallic state. [c] The total density of states for this metallic state, computed within GGA+SOC+$U$. The inset shows the band structure near ($\rm {E_F}$). [d] The variations of the bond angle $\angle$(Ir-O1-Ir)  and band width with the change in Ir-O bond length across the AIAO insulating to metallic state transition in [b], showing the increase in band width as the Ir-O bond length decreases which gives rise to metallic state.}
	\label{dos}
\end{figure*}	
\begin{table*}

\begin{tabular}{ c c c c c  c  c c c c c c c c c c c c c c c  }
\hline
\hline
Temperature &&& Shell &&&  N &&&   $\sigma^2$ (\AA$^2$) &&&   R (\AA) &&& R$_{XRD}$ (\AA) \\ \hline
\multirow{6}{*}{T=20 K} &&& Ir-O1 (1) &&& 6* &&& 0.0042(6) &&& 1.9946(230) &&& 2.015(3)\\
 &&& Ir-Ir &&& 6* &&& 0.0015(8) &&& 3.622(5) &&& 3.642(2) \\
 &&& Ir-Eu &&& 6* &&& 0.0041(3) &&& 3.628(13) &&& 3.642(2)\\
 &&& Ir-O1 (2) &&& 6* &&& 0.0011(9) &&& 3.668(15) &&& 3.891(0)\\
 &&& Ir-O1 (3) &&& 12* &&& 0.0103(5) &&& 4.339(38) &&& 4.417(4)\\ \hline
\multirow{6}{*}{T=200 K} &&& Ir-O1 (1) &&& 6* &&& 0.0023(5) &&& 1.9952(22) &&& 2.014(1)\\
 &&& Ir-Ir &&& 6* &&& 0.0024(0) &&& 3.624(5) &&& 3.644(0) \\
 &&& Ir-Eu &&& 6* &&& 0.0069(7) &&& 3.625(17) &&& 3.644(0)\\
 &&& Ir-O1 (2) &&& 6* &&& 0.0026(9) &&& 3.667(22) &&& 3.885(8)\\
 &&& Ir-O1 (3) &&& 12* &&& 0.0132(1) &&& 4.334(36) &&& 4.272(9)\\ \hline
\hline
\end{tabular}
\caption{The local structure parameters obtained from the Ir-$L_3$ EXAFS analysis at $T$ = 20 K and 200 K. $N$ and $\sigma^2$ represent the coordination number and Debye-Waller factor respectively. The interatomic distances as obtained from the synchrotron XRD refinement and the EXAFS analysis are illustrated as $R_{XRD}$ and $R$ respectively for comparison. The absolute mismatch between the experimental data and the best fit ($\Re^2$) is found to be 0.011 and 0.012 respectively.}
\label{Ir-par}
\end{table*}
\par
The experimental BR in this sample has been calculated by integrating the resonant cross section at the Ir-$L_2$ and $L_3$ edges after subtracting a step function broadened by the core-hole lifetime to emulate the single-atom absorption process. The intensities of the Ir-$L_2$ and $L_3$ absorption edges are normalized for accurate comparison. The continuum steps at the Ir-$L_2$ and $L_3$ edges are made equal to unity and half of unity respectively. The objective behind this normalization is that the ratio of occupied 2$p$$_{1/2}$ and 2$p$$_{3/2}$ states is 1:2. We have fitted the white lines using arc-tangent and Lorentzian functions and the fitting for both  Eu$_2$Ir$_2$O$_7$ and IrO$_2$ are depicted in the inset of figs.~\ref{Ir-xanes}[c] and [d]. Our calculation gives rise to a BR value $\sim$2.76 at $T$ =200 K in contrast to the existing strongly spin-orbit coupled 5$d$ iridates~\cite{Laguna-PRL,Clancy-PRB2012,Tafti-PRB,Aczel-PRB2019}, and it is not significantly higher than the statistical BR value. Such a value of BR (~2.76 at 200K), even as compared to that of elemental Ir ($\sim$3.2-3.6) ~\cite{Clancy-PRB2012}, suggests that the effective SOC strength on Ir 5$d$ is at most moderate in this compound~\cite{Laguna-PRL,Clancy-PRB2012,Tafti-PRB,Aczel-PRB2019}, but certainly far away from the atomic $jj$ coupling limit. To further verify the reliability of the value of BR, we have  calculated the BR for IrO$_2$ from the EXAFS data at 200 K in the same instrument, following the same procedure. The value of BR for IrO$_2$ was found to be 4.04 from our analysis, which closely matches with that of the previous report~\cite{Cho-JPCM}, validating our method of BR calculation. Therefore, it is likely that the combined effects of crystal distortion, Ir-O covalency, hybridization of Ir-5$d$ orbitals, intersite Ir-Ir hopping, and the Ir-5$d$ bandwidth suppress the effective SOC strength in  Eu$_2$Ir$_2$O$_7$.

\subsection {Extended x-ray absorption fine structure}
To identify the local atomic structure and its possible change across the  MIT, the temperature dependent ($T$ = 200 to 10 K) Ir-$L_3$ edge EXAFS measurements were carried out. A general impression on these collected Ir-$L_3$ EXAFS data  and the respective Fourier transforms (FTs) (fig.~\ref{Ir-exafs} [c]) leads to the fact that the overall local structure around the Ir-cation is preserved without noticeable modifications with decreasing temperature down to 10 K.
\par
The $\kappa^3$ (here $\kappa$ is the wave number of the photoelectrons)  weighted experimental EXAFS data were fitted using the multishell data refinement procedure in the range of 1-4.5 \AA~ to access the structural information around Ir. We considered the nearest neighbor Ir-O1 and next nearest neighbor Ir-Ir and Ir-Eu single scattering paths. Two other single scattering Ir-O1 paths at higher atomic distances ($\sim$3.77 and 4.45 \AA) are taken into account for better refinements. The first shell in the FT data at around 1.7 \AA~ comprises of the Ir-O1 nearest neighbor contribution and the second shell at around 3.5 \AA~ is contributed from both the Eu and Ir neighbors. As seen in fig.~\ref{Ir-exafs} [a], the larger intensity of the second shell (compared to the first shell at all the temperatures) clearly suggests predominance of the combined weightage from Ir-Eu and Ir-Ir single scattering paths relative to the nearest-neighbor Ir-O1 scattering path. The results found from our EXAFS analysis are summarized in Table~\ref{Ir-par}. It strongly refutes any Eu-Ir antisite disorder by the perfectly retained local atomic coordination geometry to the stoichiometrically expected numbers at all the temperatures. By fitting the Ir-$L_3$ EXAFS data recorded at different temperatures, we have plotted the  $T$ variation of the local correlated Ir-O1 bond length as shown in fig.~\ref{Ir-exafs} [d]. Initially, Ir-O1 distance increases as we decrease $T$. However, it starts to decrease with decreasing $T$ below $T_{MI}$ constituting a hump like feature peaking around $T_{MI}$. This $T$ variation of Ir-O1 agrees well with our XRD result (fig.~\ref{pxrd} [c]).

\subsection{Neutron powder diffraction}
Figs.~\ref{pnd1} [a] and [b] show the neutron powder diffraction patterns measured at 150 and 1.5 K, respectively, $ i.e.$, above and below the transition temperature. The lattice parameter and $x$ coordinate of the O1 (Wyckoff position 48$f$)  at 150 K  are found equal to $a$ = 10.2991(2) \AA~ and $\xi_1$ = 0.336(5). From fig. ~\ref{pnd1}[b], we can see that at 1.5 K, only fundamental reflections of the pyrochlore structure are observed, and no extra peak for magnetic ordering is detected. This is in contrast to the neutron diffraction study on Nd$_2$Ir$_2$O$_7$, where a single sharp magnetic Bragg peak appears at 3 K associated with the ordering of the rare-earth. The absence of any magnetic peak down to 1.5 K in  Eu$_2$Ir$_2$O$_7$ possibly indicate the lack of ordered moment at the  Eu-site, which is in line with the  $\mu$SR study ~\cite{zhao}. We did not observe any diffuse magnetic scattering or line broadening in our NPD data, which are  commonly seen in the magnetic rare-earth pyrochlore oxides in the  spin-ice state ~\cite{Aczel-PRB2019}.

\par
Despite the fact that no magnetic reflection is observed at 1.5 K, there is a  small but noticeable increase in the intensity of the nuclear peaks at low temperature.  The  integrated intensities of  (113) and (222) show the most significant increment (inset of fig.~\ref{pnd1} [b]). We have carefully analysed the NPD and the PXRD data, and it appears that the increment in intensity of the above mentioned nuclear peaks is significantly larger than that accounted by the thermal effects (Debye-Waller factor). This increment most likely arises from the magnetic ordering of the Ir sublattice; based on the propagation vector $k$ = (0, 0, 0). Among  four allowed magnetic structures compatible with the $Fd\bar{3}m$ symmetry (see section~\ref{mag-th}), the AIAO type model (fig.~\ref{spin} [a]) is found to be the most appropriate which is in line with the magnetic structure of other pyrochlore iridates. The magnetic Bragg positions are overlapped with the nuclear reflections, and there is a minimal increase observed at low temperature due to the high neutron absorption cross-section of Ir. The magnetic moment of Ir can not be considered a freely refinable parameter in the Rietveld refinement process due to the weak statistics of the data. In presence of crystal field and spin-orbit coupling, the effective angular momentum for the Ir$^{4+}$ state is $J_{eff}$ = 1/2 and the corresponding magnetic moment is  $\mu_{eff} = g_JJ_{eff}\mu_B$ = (2/3)(1/2) $\mu_B \sim$ 0.33 $\mu_B$. We performed our refinement of NPD data considering this effective value of Ir magnetic moment, and we observed that the data are well fitted in the range 0.28 to 0.40 $\mu_B$/Ir.  Our attempts to refine the neutron data by considering an ordered magnetic moment at Eu and Ir sites only worsens the fit. Therefore, we propose that Eu$_2$Ir$_2$O$_7$ presumably attains an AIAO magnetic structre of Ir spins.

\section{Theoretical Results}

\subsection{Magnetic structure analysis}
\label{mag-th}

We start with the analysis of the magnetic structure of Eu$_2$Ir$_2$O$_7$, where the conventional unit cell contains 8 formula units \cite{millican}. To do this, we first determine the allowed magnetic space groups for the given crystal structure of Eu$_2$Ir$_2$O$_7$ using \texttt{MAXMAGN}, as implemented in the Bilbao crystallographic server \cite{theo}. For the experimentally determined magnetic propagation vector $\vec k = (0, 0, 0)$, we find that among the several allowed magnetic space groups, only four magnetic space groups are allowed for the non-zero magnetic moment at all Ir sites. These four allowed magnetic space groups are [a] $Fd$-${3}m^\prime$ , [b] $I4_1/am^\prime d^\prime$ , [c] $I4_1^\prime /amd^\prime$, and [d]  $Imm^\prime a^\prime$~\cite{theo}. The corresponding magnetic configurations are depicted in fig. \ref{spin}.

\par
The total energies of these various symmetry allowed magnetic configurations are calculated within the GGA+SOC+U formalism and the computed values are listed in Table~\ref{energy}. The comparison of the computed total energies shows that the magnetic ground state of  Eu$_2$Ir$_2$O$_7$ corresponds to $Fd$-$3m^\prime$ magnetic space group. The spin moments at Ir$^{4+}$ ions are  arranged on a corner shared tetrahedra, and are directed either towards the center of the tetrahedra or away from it, thereby forming an AIAO magnetic structure (see fig. \ref{spin} [a]). The spin and orbital moments at the Ir site in this magnetic structure are respectively 0.40 $\mu_B$ and 0.29 $\mu_B$, indicating the presence of a moderate SOC in the system due to Ir. The corresponding energy spectrum is gapped out as can  be seen from the computed total densities of states, shown in fig. \ref{dos} [a].

\begin{table} [h]
	\caption{Energy differences and the spin moments for non-magnetic and various symmetry allowed magnetic
		configurations for Eu$_2$Ir$_2$O$_7$, within GGA+SOC+$U$ with
		$U_{\rm eff}$ = 1.5 eV. The orbital moment at the Ir site is presented within the
		parentheses.}
	\centering
	\begin{tabular}{ c c c   c c}
		\hline
		Magnetic     &  $\Delta E$ & Moment/Ir   &  Total      & Energy \\
		Space group  & (in meV/fu)  & (Orbital)   &  Mom./fu    & Gap  \\
		&              & in $\mu_B$  &  in $\mu_B$ & (eV)\\[1 ex]
		\hline
		NM          &   80  & 0   &  0   &  0 \\
		$Fd\bar{3}m^\prime$      &    0  & 0.40 (0.29)   &  0.0   & 0.215\\
		$I4_1/am^\prime d^\prime$ &   42  & 0.36 (0.37) & 2.29 & 0.0 \\
		$I4_1^\prime/amd^\prime$ & 8  &   0.11 (0.32)  &  0.0    & 0.198\\
		$Imm^\prime a^\prime$ & 81 &  0.11 (0.04)     &  3.26   & 0.0\\
		&    &  0.16 (0.06)     &         &   \\
		\hline
	\end{tabular}
	\label{energy}
\end{table}

\subsection{Role of ${\rm Ir-O1}$ bond length and bond angle $\angle$(Ir-O1-Ir) in metal-insulator transition}

To investigate the roles of Ir-O1 bond length and  $\angle$(Ir-O1-Ir) bond angle in MIT in Eu$_2$Ir$_2$O$_7$, we varied the bond length and the bond angle, and examined the  corresponding changes in the electronic and magnetic properties of Eu$_2$Ir$_2$O$_7$. The changes in the bond length and bond angle are achieved by changing the only free parameter $\xi_1$ ~\cite{millican}. Both Ir-O1 bond length and $\angle$(Ir-O1-Ir) bond angle can be tuned by artificially changing the position of O1 atom, which provides the platform to study the role of crystal geometry in the MIT.
\par
We further examined the stability of the AIAO insulating ground state as discussed in the last section, with respect to the variations of the bond lengths and bond angles. As seen from figs.~\ref{dos} [b] and [d], larger Ir-O1 bond length and smaller bond angles favor the stabilization of the AIAO magnetic state. In this region, the spin moment at the Ir site increases with the increase in Ir-O1 bond length and the simultaneous decrease in  $\angle$(Ir-O1-Ir). Interestingly, the band gap  increases accordingly, indicating that the AIAO AFM state plays an important role in the insulating state of  Eu$_2$Ir$_2$O$_7$. Our results support the previous experimental findings~\cite{ishi}, which suggest the importance of magnetic order in opening up a charge gap in the system.
\par
The situation becomes even more interesting in the regime of small Ir-O1 bond length. As the bond length decreases, the system first undergoes a transition from AIAO magnetic insulating state to a metallic state and, then, from the metallic state to a non-magnetic (NM) insulating state, as shown in fig.~\ref{dos} [b].  The density of states (DOS) as well as the band structure near the Fermi energy ($E_F$) for the metallic state are shown in fig.~\ref{dos} [c]. The metallic phase of Eu$_2$Ir$_2$O$_7$ has potential for exhibiting Weyl physics~\cite{liu,Wang2017}. In our calculation, we found several band crossings near $E_F$  (see inset of fig.~\ref{dos} [c]). The determination of the topological nature of these band crossings, however, requires further analysis of the structural symmetry, surface state properties, etc.,  and can be considered as a future direction of work.
\par

\begin{figure}[t!]
	\centering
	\includegraphics[ width= 8 cm]{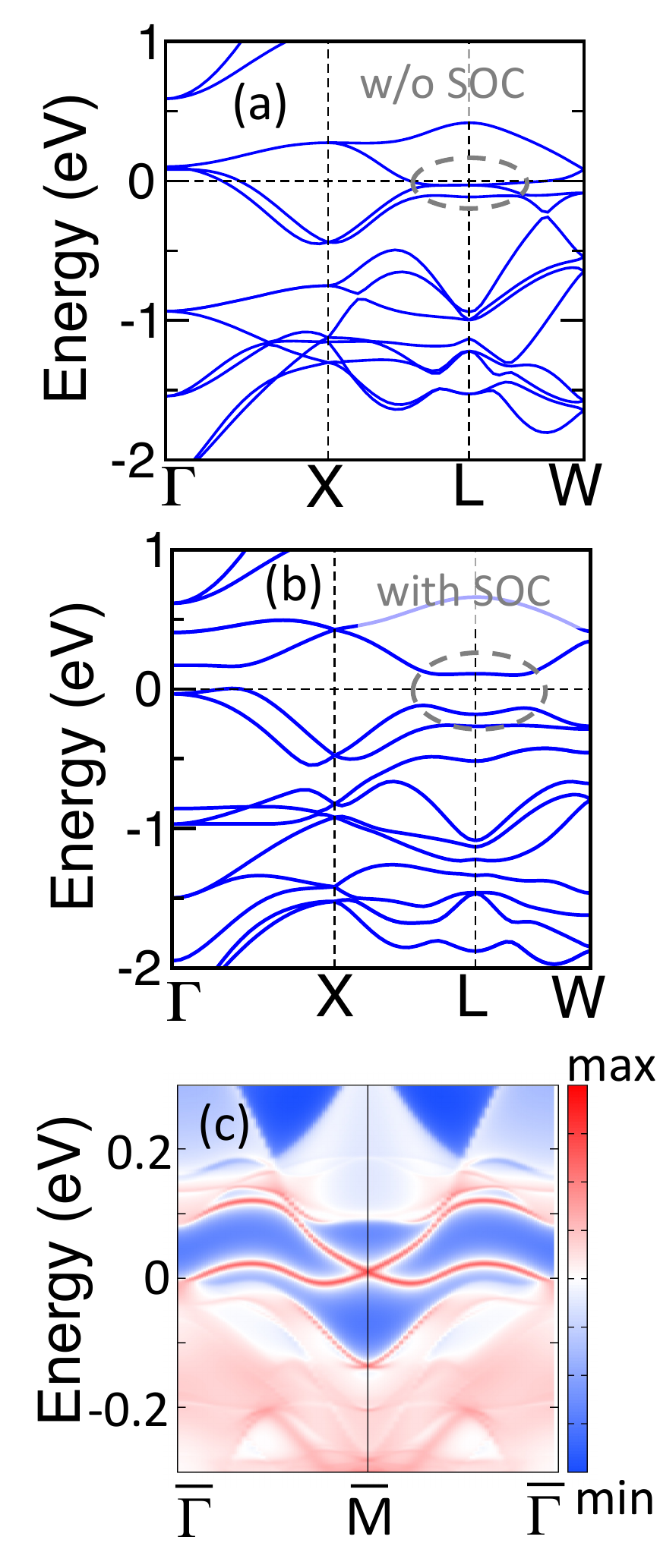}
	\caption{Bandstructure of Eu$_2$Ir$_2$O$_7$ for the small Ir-O bond length regime in [a] absence and [b] presence of spin-orbit coupling (SOC). The dashed circle in [b] indicates the band inversion in presence of SOC (see text for details). [c] The surface density of states for the (110) surface, showing the topological surface states (solid red lines) connecting the valence and conduction bands. Blue regions denote the bulk energy gap. The Fermi energy is set at zero. }
	\label{TI}
\end{figure}

	We now turn to the insulating regime of very small Ir-O bond length. The bandstructures in the regime of Ir-O bond length are shown in figs. \ref{TI}  [a] and [b] in absence and presence of spin-orbit interaction respectively. The comparison of these bandstructures indicates the presence of a band inversion when spin-orbit effects are present, suggesting a topological state for the small Ir-O bond length.  To confirm this, we investigate 
the topological property of the NM insulating state within the topological quantum chemistry formalism~\cite{TQC}. It is based on (i) elementary band representations (EBRs)~\cite{ebr} that constitutes the building blocks of bands which come out of atomic orbitals, and (ii)  Brillouin zone (BZ) compatibility relations~\cite{compatibility}, dictating all possible band connectivities in the BZ. An insulator is said to be topological if  the valence bands in that insulator can not be written as the sum of the EBRs \cite{TI}. The topological phase can, further, be classified into \enquote{strong} or \enquote{fragile}. For the \enquote{strong} topological phase, the characters of the representations at the maximal k-points (a set of high symmetric k-points for a certain symmetry of the crystal structure) can not be written as the sum of characters of the EBRs in that group. On the other hand, in the \enquote{fragile} topological phase, a set of bands can be written as a difference plus a sum in contrast to the sum (only) of EBRs.  To compute the topological phase of the NM insulating state in the small Ir-O1 bond length regime, we have used  Bilbao crystallographic server and \texttt{vasp2trace} program \cite{bilbao}. Our calculation shows that Eu$_2$Ir$_2$O$_7$ enters into a  topological phase ($Z_2 =1$) for smaller Ir-O1 bond length. 


	For further confirmation of the topologically nontrivial state, we compute the $Z_2$ invariant based on the evolution of the Wannier charge center using WannierTools \cite{Wu2018}. We compute the evolution of the Wannier charge centers of the occupied Bloch bands for the six time-reversal invariant momentum planes, viz., $k_1=0 ~{\rm{\text and}}~\pi, k_2=0 ~{\rm{\text and}}~ \pi, k_3=0 ~{\rm{\text and}}~ \pi $. The number (even or odd) of crossings between a reference line and the evolution line determines the $Z_2$ index (zero or one) for a particular plane \cite{Soluyanov2011,Gresch2017}. Our calculation shows $Z_2 =1$ for $k_i=0$ and $Z_2 =0$ for $k_i=\pi$ planes,  where $i=1-3$. We, then, compute the $Z_2$ invariant $\nu_0; (\nu_1\nu_2\nu_3)$ \cite{FuKane2007} of the crystal using the following relations,
	\begin{eqnarray} \nonumber
		\nu_0 &= &\big( Z_2|_{k_i=0} +Z_2|_{k_i=\pi} \big)~ {\rm{\text mod}}~ 2 \\ 
		\nu_i &=& Z_2|_{k_i =\pi}, ~ {\rm{\text for}}~i=1-3.
	\end{eqnarray}
	The computed $Z_2$ invariant  is 1;(000) confirming a strong topological insulating state for the small Ir-O bond length regime, consistent with our previous analysis. The computed $Z_2$ invariant also indicate the presence of a topologically protected surface state in any termination. The computed surface densities of states for the (110) surface is shown in fig. \ref{TI} [c]. The topologically nontrivial surface state connecting the valence and conduction bands becomes apparent from this figure.

\par
 Returning to the discussion of bond length variation,  the magnetic and electronic behavior of Eu$_2$Ir$_2$O$_7$ with the variation of Ir-O1 bond length and $\angle$(Ir-O1-Ir), as discussed above, may be attributed to the changes in the band width and the trigonal crystal field splitting in the system. To understand these changes in the electronic structure with the change in crystal geometry, we further studied the corresponding variations in the bare Ir-$t_{2g}$ band width (non-spin polarized band width in absence of SOC). As shown in fig.~\ref{dos} [d], the band width increases with the simultaneous decrease in Ir-O1 bond length and increase in $\angle$(Ir-O1-Ir), likely due to the larger overlap between Ir-$d$ and O-$p$ states. This increase in band width, in turn, results in a metallic ground state. We note that although DFT calculations overestimate the changes in  bond angle and the bond lengths across the MIT compared to the measured values, the obtained trend from our calculations is consistent with the experimental observations.
\par
The IrO1$_6$ octahedra  undergo trigonal compression, as a result of which Ir-t$_{2g}$ level splits into lower lying singly degenerate a$_{1g}$ and doubly degenerate e$^\pi_g$ levels.  An  estimation of the trigonal splitting is obtained using muffin-tin orbital (MTO) based N$^{th}$ order MTO (NMTO) method
~\cite{NMTO} , and the computed value is about  $\sim$ 0.22 eV.  The trigonal splitting is expected to enhance the
spin moment in the pyrochlore iridate \cite{Shinaoka}. This explains the increase in the spin moment as the Ir-O1 bond length increases. Note that in our calculation, larger Ir-O1 bond length indicates stronger deviation of $\xi_1$  from $\xi_{1c}$, where $\xi_{1c}$ = 0.3125 is the ideal value, and, therefore, represents  stronger distortion. Indeed, model calculations in an earlier work~\cite{yang2} show that trigonal crystal field in real systems is responsible for the destabilization of a topological insulating state, predicted for an ideal cubic environment~\cite{pesin}. The revival of such topological insulating state is also predicted by Yang {\it et. al}~\cite{yang2} via softening of the $q=0$ modes corresponding to trigonal distortion. Our results indicate that the topological insulating state at small Ir-O1 bond length may be an exemplary of this model, which  predicts a topological insulating state that may be achieved under high pressure.  We note that the previous experiments under high-pressure condition \cite{Tafti2012} found an insulator to metal transition in agreement with our theoretical prediction, but did not find any clear evidence of topological insulating phase (the possibility of such a state at $T \rightarrow$ 0 K was although not ruled out) as obtained in the present work. This is likely related to the very small Ir-O bond length in the predicted topological regime which is far below the Ir-O bond length obtained in the high pressure measurements \cite{Clancy-prb2016}, indicating the requirement of an even smaller Ir-O bond length to achieve the predicted topological insulting state.

\section{Discussion}
We begin our discussion by commenting on the SOC strength which, although is predicted to play an important role, has never been accurately determined in this important pyrochlore iridate. It is clear from our branching ratio calculation that effective SOC strength on Ir-5$d$ in Eu$_2$Ir$_2$O$_7$ is significantly lower than elemental Ir or that in IrO$_2$. We have discussed the possible causes such as, Ir-Ir intersite hopping or Ir 5$d$ bandwidth which could suppress the SOC.  Indeed, the BR provides us with the expectation value of the spin-orbit operator $<{\bf L.S}>$, which is certainly not the atomic SOC, but a renormalized effective SOC in the crystalline solid ~\cite{psmio-prb,Liu-PRL,Cd2Ir2O7-prb,Paramekanti-PRB,Takegami-PRB}. Moreover, the  trigonally distorted IrO1$_6$ octahedra, and the subsequent triply degenerate Ir $t_{\text{2}g}$ energy level splitting would eventually rearrange the spin-orbit-derived $J$-states, which can be associated with the lower value of effective SOC ~\cite{Cd2Ir2O7-prb,trigonal1,Pramanik-PRB2020}. This picture is supported by the previous  resonant elastic and inelastic x-ray scattering studies which estimated the energy level splitting to be 0.45 eV due to the trigonal distortion close to the typical value of atomic SOC of 0.5 eV for  Ir$^{4+}$ state~\cite{trigonal1}.

\par
A key experimental observation of our study  is the change in the Ir-O1 bond length and the Ir-O1-Ir bond angle with $T$. $\angle$(Ir-O1-Ir) shows a sharp drop at $T_{MI}$ on cooling, while the temperature variation of Ir-O1 bond distance reveals a peak like feature. Interestingly, these anomalies coincide with the AIAO long-range AFM ordering. Therefore the structural anomalies occurring at $T_N$ can  be ascribed to the exchange striction mechanism. A competition between the Heisenberg exchange energy ($\mathcal{J}(R_{ij}) s_i.s_j$) between the spins {\bf $s_i$} and {\bf $s_j$}  and the elastic energy is responsible for the structural relaxation  of the magnetically coupled ions in the lattice. A balance is reached when the interionic distance $R_{ij}$ changes suitably to compensate  the additional magnetic energy due to the  long-range magnetic ordering~\cite{kittel,baidya,sagayama,lines}. The anomaly in  $\angle$(Ir-O1-Ir) at $T_N$ corroborates well with  the change observed in the Ir-O-Ir bending modes in Raman spectroscopy~\cite{uda}.

\par
A decrease in the $\angle$(Ir-O1-Ir) will eventually reduce the hybridization of Ir-5$d$ orbitals with O-2$p$. The 5$d$ bandwidth of the transition metal is given by~\cite{khomskii}:

\begin{equation}
t_{dd}^{eff} = t_{pd}^2/\Delta_{CT}
\end{equation}

where, $t_{dd}^{eff}$ is the effective intersite electron hopping strength among the $d$ levels of the transition metal, $t_{pd}$ is the intersite electron hopping between $d$ state of the transition metal and the $p$ state of oxygen, and $\Delta_{CT}$ is the charge transfer energy required to transfer an electron from the $p$ level of oxygen to the transition metal's $d$ level.  A weakening of $t_{pd}$ through a decrease in the Ir-O1-Ir bond angle will reduce the electron hopping among the neighboring Ir 5$d$ orbitals subsequently lowering the orbital-overlap integral, and  the hybridization of Ir-5$d$ orbitals leading to  an insulating state below $T_{MI}$. The reduced electron hopping among the Ir 5$d$ orbitals strengthens the superexchange interactions between the neighboring Ir ions via O1. This  will  favor the long-range AFM ordering in this compound at $T_{MI}$. In the  $R_2$Ir$_2$O$_7$ family, $T_{MI}$ decreases and  $\angle$(Ir-O1-Ir) increases with the increase of the radius of $R$-cation~\cite{matsuhira}. This indicates that a lower value of $\angle$(Ir-O1-Ir) favors the insulating state. This is consistent with our present finding, where we find a lowering trend of $\angle$(Ir-O1-Ir) around $T_{MI}$.

\par
Our  neutron  analysis  possibly precludes the presence of any ordered moment at the Eu site, and shows a magnetic moment value ranging between 0.28 and 0.4 $\mu_B$ moment at the Ir site from the Rietveld refinement of the NPD data. At this point it is worth mentioning the ordered moment values of few other pyrochlore iridates having similar AIAO type magnetic structure.  We note here that the upper limit of the Ir moment obtained from the neutron diffraction study on Nd$_2$Ir$_2$O$_7$ is $\sim$ 0.2 $\mu_B$ when the interaction between Nd and Ir is ferromagnetic. The neutron diffraction measurement on Y$_2$Ir$_2$O$_7$ put an upper bound for the ordered Ir moment below $T_{MI}$ of $\sim$ 0.5 $\mu_B$ for the $k$ = (0, 0, 0) magnetic structure~\cite{shapiro}. 

\par
By using symmetry arguments and explicit DFT calculations we demonstrate independently that the AIAO AFM structure of Ir$^{4+}$ tetrahedra constitutes the magnetic ground state of Eu$_2$Ir$_2$O$_7$, which is in agreement with our neutron measurement. Importantly, our calculations clearly show that the Ir-O1 bond length and $\angle$(Ir-O1-Ir) play crucial roles in the metal-insulator transition, in particular stabilizing the AIAO magnetic structure, which, in turn, stabilizes the insulating state. Thus the AIAO magnetic order itself is important for the insulating state in  Eu$_2$Ir$_2$O$_7$.
\par
Anomalies in Ir-O1 bond length and $\angle$(Ir-O1-Ir) across the MIT is further examined by the first principle calculations at $T$=0 K. For smaller values of $\angle$(Ir-O1-Ir), the DFT calculations indicate the emergence of an insulating state which is in line with our experimental data. Another intriguing aspect of the present  DFT calculations is the  emergence of a  topological insulating state in the small Ir-O1 bond length regime, which may be achieved in real experimental scenario by applying external pressure to the system.

\section{Conclusions}
Our exhaustive study combining synchrotron based x-ray diffraction and x-ray absorption spectroscopy with high resolution neutron diffraction and electronic structure calculations show the underlying role of the  lattice and magnetic structures in determining the electronic properties of Eu$_2$Ir$_2$O$_7$. Importantly our DFT calculations predict the possibility of a topological insulating phase of Eu$_2$Ir$_2$O$_7$  for a short Ir-O1 bond length, which implies future high pressure studies might reveal a rich and yet unexplored phase in this system.

 \section{Acknowledgment}
 The work is supported by the financial grant from UKIERI project [Grant No.
 DST/INT/UK/P-132/2016]. MD would like to thank CSIR, India for her research fellowship [File No. 09/080(1027)/2016-EMR-I]. The experiments  at Petra-III and ISIS  are supported by the  India@DESY Partnership and  India-RAL collaborative project (SR/NM/Z-07/2015) respectively. IDG thanks SERB-India and TRC (DST) for the financial support. Computational resources were provided by Indian Association for the Cultivation of Science (\texttt{Monami2} cluster).  SB thanks ETH Zurich for financial support. We acknowledge Deutsches Elektronen-Synchrotron (Hamburg, Germany), a member of the Helmholtz Association HGF, for the provision of experimental facilities. Parts of this research were carried out at PETRA III and we would like to thank Ann-Christin Dippel, Oleh Ivashko, Philipp Glaevecke, and Martin v. Zimmermann for assistance in using P21.1 beamline (proposal no. I-20200448). The Diamond Light Source (UK) is acknowledged for time on beamline B-18 under Proposal SP17752-1.

\end{document}